# Querying Schemas With Access Restrictions

Michael Benedikt
Oxford University, UK
michael.benedikt@cs.ox.ac.uk

Pierre Bourhis
Oxford University, UK
pierre.bourhis@cs.ox.ac.uk

Clemens Ley
EPFL, Switzerland
ley.clemens@gmail.com

## ABSTRACT

We study verification of systems whose transitions consist of *accesses to a Web-based data-source*. An access is a lookup on a relation within a relational database, fixing values for a set of positions in the relation. For example, a transition can represent access to a Web form, where the user is restricted to filling in values for a particular set of fields. We look at verifying properties of a schema describing the possible accesses of such a system. We present a language where one can describe the properties of an access path, and also specify additional restrictions on accesses that are enforced by the schema. Our main property language, AccLTL, is based on a first-order extension of linear-time temporal logic, interpreting access paths as sequences of relational structures. We also present a lower-level automaton model, A-automata, which AccLTL specifications can compile into. We show that AccLTL and A-automata can express static analysis problems related to "querying with limited access patterns" that have been studied in the database literature in the past, such as whether an access is relevant to answering a query, and whether two queries are equivalent in the accessible data they can return. We prove decidability and complexity results for several restrictions and variants of AccLTL, and explain which properties of paths can be expressed in each restriction.

## 1. INTRODUCTION

Many data sources do not expose either a bulk export facility or a query-based interface, enforcing instead many restrictions on the way data is accessed. For example, access to data may only be possible through Web forms, which require bindings for particular fields in the relation [16, 4]. Querying with limited access patterns also arises in other middleware contexts (e.g. federated access to data in Web services) as well as in construction of query interfaces on top of pre-determined indexed accesses [20]. For example, a Web telephone directory might allow several Web forms that serve as *access methods* to the underlying data. It may have an access method $AcM_1$ accessing a relation

$$Mobile\#(\underline{name}, postcode, street, phoneno),$$

where $AcM_1$ allows one to enter a mobile phone customer's name (the underlined field) and access the corresponding set of tuples containing a postal code, mobile phone number and street name. The same site might have an access method $AcM_2$ on relation

$$Address(\underline{street}, \underline{postcode}, name, houseno)$$

allowing the user to enter a street name and postcode, returning all corresponding resident names and housenumbers. Formally an access method consists of a relation and a collection of *input positions*: for $AcM_1$, position 1 is the sole input position, while for $AcM_2$ the first two positions are input. An *access* consists of an access method plus a binding for the input positions – for example putting "Smith" into method $AcM_1$ is an access. The *response* to an access is a collection of tuples for the relation that agree with the binding given in the access. A schema of this sort defines a collection of *access paths*: sequences consisting of accesses and their responses.

The impact of "limited access patterns" has thus been the subject of much study in the past decade. It is known that in the presence of limited access patterns, there may be no access path that completely answers the query, and there may also be many quite distinct paths. For example, the query $Address(X, Y, \text{"Jones"}, Z)$ asking for the address of Jones is not answerable using the access methods $AcM_1$ and $AcM_2$ above. There are certainly many ways to obtain the maximal answers: one could begin by obtaining all the street names and postcodes associated with Jones in the Mobile# table, entering these into the Address table to see if they match Jones, then taking all the new resident names we have discovered and repeating the process, until a fixedpoint is reached. If, however, Jones does not occur as a name in Mobile#, then this process will not yield Jones' tuple in Address. In general it is known [15] that for any conjunctive query one can construct (in linear time) a Datalog program that produces the maximal answers to a query under access patterns: the program simply tries all possible valid accesses on the database, as in the brute-force algorithm above.

In the absence of a complete plan, how can we determine which strategy for making accesses is best? Recent works [4, 3] have proposed optimizing recursive plans, using access pattern analysis to determine that *certain kinds of accesses can not extend to a useful path*. An example is the work in [3] which proposes limiting the number of accesses to be explored by determining that some accesses are not "relevant" to a query. An access is *long term relevant* if there is an access path that begins with the access and uncovers a new query result, where the removal of the access results in the new result not being discovered. [3] gives the complexity of determining relevance for a number of query languages.

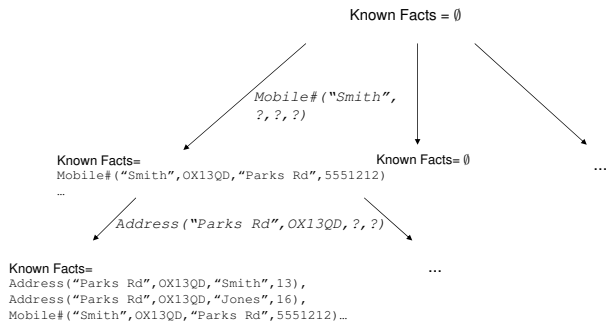

**Figure 1: Tree of possible paths associated with a schema**

Long term relevance is only one property that can be used to measure the value of making a particular access – for example we may want to know whether there is an access that reveals several values in the query result. Furthermore, "limited access patterns" represent only one possible restriction that limits the possible access paths through a web interface. Many other restrictions may be enforced, e.g:

- Restrictions that follow from *integrity constraints on the data*: e.g. a mobile phone customer name will not (arguably) overlap with a street name. Thus in an iterative process for answering the query given above, we should not bother to make accesses to the Mobile# table using street names we have acquired earlier in the process. It is also easy to see that key constraints, and more generally *functional dependencies*, can play a crucial role in determining whether an access is relevant.
- *Access order restrictions*: e.g. before making any access to Mobile#, the interface may require a web user to have made at least one access to Address.
- *Dataflow restrictions*; before performing an access to Mobile# on a name, the web user must have received that name as a response to a call to Address.

Ideally, a query processor should be able to inspect an access and determine whether it is a good candidate for use, where the assumptions on the paths as well as the notion of "good candidate" could be specified on a per-application basis. In this paper we look for a general solution to specifying and determining which accesses are promising: *a language for querying the access paths that can occur in a schema*. We show that every schema can be associated with a labelled transition system (LTS), with transitions for each access and nodes for each "revealed instance" (information known after a set of accesses). A fragment of the LTS for the schema with access methods $AcM_1$ and $AcM_2$ is given in Figure 1. Paths through the LTS represent possible access/response sequences of the Web-based datasource. There are infinitely many paths – in fact every access could have many possible responses. But the access restrictions in the schema place limitations on what paths one can find in the LTS. We can then identify a "query on access paths" with a query over this transition system. This work will provide a language that allows the user to ask whether a given kind of path through instances of the schema is possible: e.g. is there a path that leads to an instance where a given conjunctive query holds, but where the path never uses access $AcM_1$? Is there a path that satisfies a given set of additional dataflow, access order restrictions, or data integrity constraints?

Paths are often queried with *temporal logic* [13]. We will look at natural variations of First-Order Linear Temporal Logic (FOLTL) for querying access paths. We look at a family of languages denoted AccLTL($L$) ("Access LTL"), parameterized by a fragment $L$ of relational calculus. It has a two-tiered structure: at the top level are temporal operators ("eventually", "until") that describe navigation between transitions in a path. The second tier looks at a particular transition, where we have first-order (i.e. relational calculus) queries that can ask whether the transitions satisfy a given property described in $L$. The relational vocabulary we consider for the "lower tier" will allow us to describe transitions given by accesses; it allows us to refer to the bindings of the access, the access method used, and the pre- and post-access versions of each schema relation. Consider the following AccLTL sentence:

$$(\neg \exists n\, \exists p\, \exists s\, \exists ph\ \text{Mobile\#}_{\text{pre}}(n,p,s,ph))\ \mathsf{U}$$
$$(\exists n\ \text{IsBind}_{AcM_1}(n) \wedge \exists s\, \exists p\, \exists h\ \text{Address}_{\text{pre}}(s,p,n,h))$$

The relational query prior to the "until" symbol $\mathsf{U}$ states that there are no entries in Mobile#$_{\text{pre}}$ – the Mobile# table prior to the access. The query after the until symbol $\mathsf{U}$ states that an access was done with method $AcM_1$ and binding $n$, where value $n$ appeared in the Address table prior to the access. Hence this "meta query" returns the set of access paths which have no entries revealed in relation Mobile# until an access AC is performed, where AC has method $AcM_1$ and uses a name that already exists in the Address table. In this work we will not be interested in returning all paths satisfying a query (there are generally infinitely many). We will check whether there is a path satisfying a given specification. This is a question of *satisfiability* for our path query language. We may also want to check that *every* path through the system is of a certain form; this is the *validity problem* for the language – bounds for validity will follow from our results on satisfiability.

We denote the logic containing the above sentence by AccLTL($\text{FO}^{\exists+}_{\text{Acc}}$), where $\text{FO}^{\exists+}_{\text{Acc}}$ is the collection of positive existential queries over a signature consisting of: the access methods, bindings, and the pre- and post- access version of each relation used in a transition. AccLTL($\text{FO}^{\exists+}_{\text{Acc}}$) can express a wide variety of properties. Unfortunately we show that satisfiability for the logic is undecidable. However, we show that a rich sublanguage of AccLTL($\text{FO}^{\exists+}_{\text{Acc}}$), denoted AccLTL$^+$, has a decidable satisfiability problem. In AccLTL$^+$ the formulas involving the bindings only occur positively. We give bounds on the complexity of this fragment, using a novel technique of *reduction to containment problems for Datalog*. We then look at the exact complexity of smaller language fragments, and show that the complexity can go much lower – e.g. within the polynomial hierarchy. The main thing we give up in these languages is the ability to express dataflow restrictions. We also study the complexity and expressiveness of extensions of the languages with inequalities and with branching time operators. In summary, our contributions are:

- We present the first query language for reasoning about the possible paths of accesses and responses that may appear in a Web form or other limited-access datasource.
- We show that combining a natural decidable logic for temporal data (LTL) with conjunctive queries gives an undecidable path query language.
- We show that by restricting to queries that are "binding positive", we get a decidable path query language. In the process we introduce a new automaton model that corresponds to a process repeatedly querying a Web data source. We show that analysis of these "access automata" can be performed via reduction to (decidable) Datalog containment problems. The automaton and logic specification languages are powerful enough to express a rich set of data integrity constraints, access order restrictions, and data flow restrictions.
- We show that the complexity of the logic can be decreased drastically by restricting the ability to express properties of the bindings that occur in accesses. The resulting language can still express important access order and data integrity restrictions, but no dataflow restrictions.
- We determine the impact adding inequalities to the relational query language, and of adding branching operators, both in terms of expressing critical properties of accesses and on complexity of verification.

**Organization:** Section 2 gives the basic definitions related to access patterns, along with our family of languages AccLTL($L$). Section 3 gives our results about the full language AccLTL(FO$_{\text{Acc}}^{\exists+}$) while Section 4 deals with AccLTL$^+$ and its restrictions. Section 5 discusses extensions of AccLTL$^+$. Section 6 gives conclusions and overviews related work. Most proofs are deferred to the full paper.

## 2. DEFINITIONS

**Schemas and paths through a schema.** Let Types be some fixed set of datatypes, including at least the integers and booleans. Our schemas extend traditional relational schemas under the "unnamed perspective" [1]. A schema Sch includes a set of relations $\{S_1 \ldots S_n\}$, with each $S_i$ associated with a function from $\{1 \ldots n_i\}$, where $n_i$ is the arity of $S_i$, to Types. We refer to the set $\{1 \ldots n_i\}$ as the *positions* of $S_i$ and the output of the function as the *domain* of the $j^{th}$ position. An *instance* $I$ for the schema consists of a finite collection $I(S_i)$ of tuples for each relation $S_i$, where a tuple is a function from the positions of $S_i$ to the corresponding domain.

A schema will also have a collection of *access methods*, where each method AcM is associated with a relation $S_i$ and a collection of *input positions* Inp(AcM). Informally, each access method allows one to input a tuple of values for Inp(AcM) and get as a result a set of matching tuples.

An *access* consists of an access method and a binding – a mapping taking the input positions of the method to their domains. A *boolean access* is one where the access method has as inputs every position of the relation – it is thus a membership test. We will use an intuitive notation for accesses, often omitting the access method. Mobile#("Jones", ?, ?, ?) is an access to relation Mobile# asking for all phone number information for people named "Jones". A boolean access is Mobile#("Jones","OX13QD","Parks Rd", 23)?, where we add the ? to make clear it is an access.

Given an access (AcM,$\bar{b}$), a *well-formed output* for AcM (on instance I) is any set of tuples r in I in the relation of AcM that is compatible with $\bar{b}$ on the input positions. We also refer to this as a *well-formed response*.

A sequence $((\text{AcM}_1, \bar{b}_1), r_1), \ldots, ((\text{AcM}_n, \bar{b}_n), r_n)$ of accesses and well-formed responses for some instance I is an *access path for the instance* I. We also refer to any sequence of accesses and responses as an access path (without reference to any instance). Note that every such sequence is an access path for *some* instance – the instance containing all returned tuples. Given an access path $p$ and an initial instance $I_0$ the *configuration returned by $p$ on $I_0$*, Conf($p, I_0$) is the instance where relation $S_i$ contains $I_0(S_i)$ unioned with all tuples returned by any access to $S_i$ in $p$. When $I_0$ is empty or understood from context we refer to the instance resulting from $p$, or Conf($p$).

As mentioned in the introduction, one is not interested in arbitrary paths, but those satisfying additional "sanity properties". We allow our schemas to prescribe some common additional properties of access methods, while additional restrictions can be expressed in the logics. The weakest property we consider here is called *idempotence*: an access path is idempotent if whenever the path repeats the same access, it obtains the same results. This corresponds to the requirement that accesses are deterministic. A stronger property is that accesses are *exact*: an access path is exact on an instance I if for every access (AcM,$\bar{b}$), the corresponding response $R$ contains exactly the tuples in the relation of AcM which agree with $\bar{b}$ on the input positions. An access path is exact if it is exact for some input instance. Put another way, an exact access path is one that contains sound and complete views of the input data for all accesses made. Most web sources are not expected to be exact – an online music site will generally not contain information about all online music. However, some forms may be known to have canonical information – e.g. a web form accessing data from a trusted government agency. We allow situations which mix exact and non-exact accesses. In general, a schema may say that some access methods are exact, some are idempotent, and some are neither. Given a set of access methods $S$, we say that an access path is $S$-exact if there is an instance I such that the path is exact for all accesses with methods in $S$, and similarly talk about $S$-idempotence.

Finally, we often do not want paths in which values for access method inputs are "guessed", but are only interested in paths where the input to an access method is a value already known. Given an instance $I_0$ (representing the "initially known information") an access path $p = a_1, r_1 \ldots$ is *grounded in $I_0$* if every value in a binding $a_i$ occurs either in $I_0$ or in a response from some $a_j$ with $j < i$. Groundedness is a special kind of dataflow restriction – our largest logics will be able to specify groundedness, along with more specialized dataflow restrictions, but we allow them also to be imposed in the schema.

A *labelled transition system* (LTS) is of the from (No, L, T) where No is a collection of nodes, L is a collection of edge labels, and T is a collection of transitions — elements of No × L × No. With any schema and initial instance $I_0$ we can associate a labelled transition system where the nodes are all the instances containing $I_0$ as a subinstance, the labels are all the accesses, and there is a transition (I, AC, I′) whenever there is some response $r$ to AC such that $\text{Conf}((\text{AC}, r), I) = I'$. We can also consider the restricted LTS where we only allow paths with transitions $(I, \text{AC}, I')$ in which the access AC is grounded at I, only paths that are idempotent, or only paths that are exact for a given subset of the access methods.

**Logics for querying access paths.** To query paths it is natural to use Linear Temporal Logic (LTL) [13]. LTL formulas define positions within a path. In Propositional LTL, the positions within paths are associated with a propositional model over some set of propositions, and one can then build up formulas from the propositions using the modal operators, $\mathsf{S}$ (since), $\mathsf{U}$ (until), $\mathsf{X}^{-1}$ (previously), $\mathsf{X}$ (next), and $\mathsf{F}$ (eventually). For example $\mathsf{F}(Q \wedge \mathsf{X} P)$ holds on positions $i$ in a path $p$ that come before some position $j$ such that proposition $Q$ holds at $j$ and proposition $P$ holds on position $j+1$. We want to extend LTL to deal with access paths, which are not just a sequence of propositional structures. Each position in an access path consists of an access and its response; the corresponding path through the LTS defined above consists of transitions $t_1 \ldots t_n$, where a transition $t_i$ is of the form $(I_i, (\text{AcM}_i, \bar{b}_i), I_{i+1})$. There is obviously a one-to-one correspondence between access paths and LTS paths as above, and we will often identify them. Since the positions carry with them a relational structure, we will use a variant of First Order Linear Temporal Logic (FOLTL) [13], which allows the use of first-order quantifiers and variables along with modal ones. We will deal here with a variant of FOLTL in which first-order sentences describing properties of positions can be nested inside temporal operators, but not vice versa.

The embedded FO formulas have the ability to constrain the instance before the access as well as afterwards. Hence, for a given vocabulary Sch, we will consider formulas over the relational vocabulary $\text{Sch}_{\text{Acc}}$ consisting of two copies $R_{\text{pre}}, R_{\text{post}}$ of each schema relation $R \in \text{Sch}$. In addition $\text{Sch}_{\text{Acc}}$ contains predicates $\text{IsBind}_{\text{AcM}}$ for each access method AcM in Sch. The arity of $\text{IsBind}_{\text{AcM}}$ is the number of input positions of AcM. An LTS path $t_1 \ldots t_n$ is associated with a sequence of $\text{Sch}_{\text{Acc}}$ structures, where the $i^{\text{th}}$ structure $M(t_i)$, corresponding to $t_i = (I_i, (\text{AcM}_i, \bar{b}_i), I_{i+1})$ interprets each predicate $R_{\text{pre}}$ using the interpretation of $R$ in $I_i$, each predicate $R_{\text{post}}$ as the interpretation of $R$ in $I_{i+1}$. The predicate $\text{IsBind}_{\text{AcM}_i}$ holds of exactly the tuple $\bar{b}_i$ while all other predicates $\text{IsBind}_{\text{AcM}}$ are empty.

We now introduce Access Linear Temporal Logic (AccLTL for short), our main specification formalism.

DEFINITION 2.1. *Let $L$ be a subset of first-order logic over $\text{Sch}_{\text{Acc}}$. The logic $\text{AccLTL}(L)$ has as atomic formulas every sentence of $L$, and is built up by the usual LTL constructors:*

$$\neg \varphi \mid \varphi \vee \varphi \mid \varphi \wedge \varphi \mid \mathsf{X} \varphi \mid \varphi \mathsf{U} \varphi$$

The semantics of $\text{AccLTL}(L)$ is given by the relation $(p, i) \vDash \varphi$, where $p = t_1 \ldots t_n$ is an LTS path and $i \leq n$. It combines the standard semantics of $L$ formulas with the usual rules for the constructors of LTL: 1. $(p, i) \vDash \varphi$ iff $\varphi \in L$ and $M(t_i)$ satisfies $\varphi$ in the usual sense of first-order logic. 2. $(p, i) \vDash \neg \varphi$ iff $(p, i) \nvDash \varphi$. 3. $(p, i) \vDash \mathsf{X} \varphi$ iff $(p, i+1) \vDash \varphi$. 4. $(p, i) \vDash \varphi \mathsf{U} \psi$ iff there exists $j \geq i$ $(p, j) \vDash \psi$ and $\forall i \leq k < j, (p, i) \vDash \varphi$. 5. $(p, i) \vDash \varphi \vee \psi$ iff $(p, i) \vDash \varphi$ or $(p, i) \vDash \psi$.

In the rest of the paper, we make use of the temporal operators $\mathsf{G}$ ("globally") and $\mathsf{F}$ ("eventually"). These operators can be expressed using $\mathsf{X}$ and $\mathsf{U}$ as usual in LTL. The *language* of a formula $\varphi$ is the set of paths $p$ such that $(p, 1) \vDash \varphi$.

Our main language of interest is $\text{AccLTL}(\text{FO}_{\text{Acc}}^{\exists +})$, where $\text{FO}_{\text{Acc}}^{\exists +}$ consists of all positive existential FO sentences over the signature $\text{Sch}_{\text{Acc}}$.

**Example 2.2** [3, 5] study query containment under (in our terminology, grounded) access patterns. Query $Q_1$ is contained in $Q_2$ relative to a schema with access patterns means that for every grounded access path $p$, if the configuration resulting from $p$ satisfies $Q_1$, then it also satisfies $Q_2$. Informally, the facts about $Q_1$ that we can determine given the schema restrictions are contained in the facts we can determine about $Q_2$. Using a containment algorithm, one can perform query minimization in the presence of access restrictions.

In [5] containment under access restrictions is shown to be decidable for conjunctive queries, while [3] studies the complexity of the problem. One can see that $Q_1$ is contained in $Q_2$ under grounded access patterns iff the following $\text{AccLTL}(\text{FO}_{\text{Acc}}^{\exists +})$ formula is a validity (over grounded paths):

$$\mathsf{G} \neg \left( Q_1^{\text{pre}} \wedge \neg Q_2^{\text{pre}} \right)$$

Here $Q_i^{\text{pre}}$ is obtained from $Q_i$ by replacing each schema predicate $S$ by $S_{\text{pre}}$ (one could as easily use $S_{\text{post}}$). We will show that containment under grounded access patterns can be expressed in a restricted fragment of $\text{AccLTL}(\text{FO}_{\text{Acc}}^{\exists +})$, as well as in an automaton-based specification formalism where validity relative to grounded access paths is decidable in 2EXPTIME. Our results will thus give tight bounds for containment under grounded access patterns.

**Example 2.3** A boolean access $\text{AC}_1$ is said to be *long term relevant* [3] (LTR) for a query $Q$ on an initial instance $I_0$ if there is an access path $p = \text{AC}_1, r_1 \text{AC}_2, r_2 \ldots$ such that the configuration I resulting from applying $p$ to $I_0$ satisfies $Q$, and the configuration resulting from the path with $\text{AC}_1$ dropped (i.e. $\text{AC}_2, r_2 \ldots$) leads to a configuration where $Q$ does not hold. In the terminology of [3] we say it is *LTR under grounded accesses* if there is a grounded access path satisfying the above.

This property can be expressed in $\text{AccLTL}(\text{FO}_{\text{Acc}}^{\exists +})$ in the following sense: for each $I_0, \text{AC}_1 = (\text{AcM}_1, \bar{b}_1)$, and $Q$ there is an $\text{AccLTL}(\text{FO}_{\text{Acc}}^{\exists +})$ formula $\varphi$ which is satisfiable iff $\text{AC}_1$ is LTR. Below we give the formula for $I_0$ being the empty instance:

$$\mathsf{F} \left( \neg Q^{\text{pre}} \wedge \text{IsBind}_{\text{AcM}_1}(\bar{b}_1) \wedge Q^{\text{post}} \right)$$

The formula checks that there is a path $p$ and a response $r_1$ to $\text{AC}_1$, such that $Q$ holds after $p$ but not after $p, \text{AC}_1, r_1$. But for a boolean access $\text{AC}_1$, the instance after $p, \text{AC}_1, r_1$ is the same as the one after $\text{AC}_1, r_1, p$.

As mentioned in the introduction, we often want additional data integrity restrictions to hold on the path. In $\text{AccLTL}(\text{FO}_{\text{Acc}}^{\exists +})$, we can add on many data integrity restrictions, such as the disjointness of names from streets, which would be expressed by a conjunction of several formulas, including:

$$\mathsf{G}(\neg \exists n \, \exists p \, \exists s \, \exists ph \, \exists hn \, \exists n' \, \exists pc \; \text{Mobile\#}_{\text{pre}}(n, p, s, ph)$$
$$\wedge \, \text{Address}_{\text{pre}}(n, pc, n', hn))$$

Similarly we can add access order restrictions and dataflow restrictions. For example, the following would restrict to paths in which names input to Mobile# must have appeared previously in Address:

$$\mathsf{G}((\exists n\ \mathrm{IsBind}_{\mathrm{AcM}_1}(n)) \to$$
$$\exists n\, \exists s\, \exists hn\, \exists pc\ \mathrm{IsBind}_{\mathrm{AcM}_1}(n) \land \mathrm{Address}_{\mathrm{pre}}(s, pc, n, hn))$$

**Example 2.4** (Data integrity restrictions, continued) Let Sch be a schema that includes, in addition to the access methods, a set of functional dependencies $d_i = R^i : pos_i \to a_i$, where $pos_i$ are positions of $R^i$ and $a_i$ is a position of $R^i$. We say that an access AcM is long-term relevant for $Q$ under Sch if there is an instance $I \supseteq I_0$ satisfying all the FDs and an access path that reveals $Q$ to be true, as in Example 2.3, but where each response returns only tuples in I.

This can be expressed in $\mathrm{AccLTL}(L_\exists^{\neq})$, where $L_\exists^{\neq}$ is the set of conjunctive queries with inequalities.

$$\mathsf{F}\left(\neg Q^{\mathrm{pre}} \land \mathrm{IsBind}_{\mathrm{AcM}}(\bar{b}_1) \land Q^{\mathrm{post}}\right) \land$$
$$\bigwedge_i \neg \mathsf{F}[\exists \vec{y}\, \vec{y}'\ R^i_{\mathrm{pre}}(\vec{y}) \land R^i_{\mathrm{pre}}(\vec{y}') \land$$
$$\bigwedge_{k \in pos_i} y_k = y'_k \land y_{a_i} \neq y'_{a_i}]$$

where $Q^{\mathrm{pre}}$ and $Q^{\mathrm{post}}$ are defined as in the previous example. We will look at languages with inequalities in Section 5.

**Basic Computational Problems.** The basic problem we consider is satisfiability of a sentence $\varphi$, which by default means that there is some access path $p$ such that $(p, 1) \models \varphi$. We will also consider satisfiability over grounded, idempotent, and (S-) exact paths.

## 3. AN EXPRESSIVE LANGUAGE FOR ACCESS RESTRICTIONS

Since satisfiability for first-order logic is undecidable, it is clear that AccLTL(FO) has an undecidable satisfiability problem. Our first main result is that the same holds even when first-order formulas are restricted to be existential.

**Theorem 3.1.** *Satisfiability of* $\mathrm{AccLTL}(\mathrm{FO}^{\exists+}_{\mathrm{Acc}})$ *is undecidable.*

This is surprising, in that $\mathrm{AccLTL}(\mathrm{FO}^{\exists+}_{\mathrm{Acc}})$ formulas deal with a fixed set of existential sentences on the configuration, and as a path progresses these queries can only move from false to true as more tuples are exposed by accesses.

The proof works by reducing the problem of determining whether a collection $\Gamma$ of functional dependencies (fds) and inclusion dependencies (ids) implies another functional dependency $\sigma$. Since this problem is known to be undecidable [6], it suffices to reduce it to unsatisfiability of an $\mathrm{AccLTL}(\mathrm{FO}^{\exists+}_{\mathrm{Acc}})$ formula.

The difficulty here is that functional dependencies seem to require negation inside a universal quantification, while inclusion dependencies require quantifier alternation – in $\mathrm{AccLTL}(\mathrm{FO}^{\exists+}_{\mathrm{Acc}})$ we have only boolean combinations of positive formula. We now explain the main idea involved in bridging this gap, which will also be used in later undecidability arguments (Theorem 5.2). The schema for our accesses includes a successor relation of a total order over the tuples of each relation in $\Gamma \cup \{\sigma\}$. The successor relation is "created" via accesses – that is, we perform accesses that reveal associations between a tuple and its successor. For each relation $R$ mentioned in $\Gamma \cup \{\sigma\}$ we also have relations $\mathrm{Beg}(R)$ and $\mathrm{End}(R)$. Our formula will enforce that these contain the first and the last tuples in the total order, respectively, by asserting the existence of additional accesses to these relations that reveal the first and last tuple. After all the relations are filled, the satisfaction of the different fd's and id's in $\Gamma$ and the failure of $\sigma$ are verified. The satisfaction of the dependencies makes use of the successor relation, and we explain the idea for FDs. We verify a dependency for one tuple at a time, iterating on the tuples according to the order. We will use a new predicate $\mathrm{Chk}^{\mathrm{FD}}(R)$ whose arity is twice the arity of $R$. This predicate will have a boolean access. $\mathrm{Chk}^{\mathrm{FD}}(R)(\vec{t}, \vec{t}')$ holding at some instance indicates that $\vec{t}, \vec{t}'$ is in accordance with the FDs on $R$. This will be done in a "nested loop" (a pair of nested "untils" in the logic) in which we iterate first over tuples $\vec{t}$, then over tuples $\vec{t}'$, accessing them progressively within $\mathrm{Chk}^{\mathrm{FD}}(R)$. At every access, we check whether the FD is satisfied, and if it is we continue the iteration.

## 4. VERIFIABLE SPECIFICATIONS: THE POSITIVE TRANSITION SUBLANGUAGE

The undecidability proof of $\mathrm{AccLTL}(\mathrm{FO}^{\exists+}_{\mathrm{Acc}})$ makes use of the ability of the logic to enforce that an access is made to a binding that does *not* satisfy a certain relation. We now consider a restriction of $\mathrm{AccLTL}(\mathrm{FO}^{\exists+}_{\mathrm{Acc}})$ which adds an additional monotonicity condition. A $\mathrm{AccLTL}(\mathrm{FO}^{\exists+}_{\mathrm{Acc}})$ formula $\varphi$ is *binding-positive* if every atom of the form $\mathrm{IsBind}(\vec{w})$ occurs only positively in $\varphi$ – that is, under an even number of negations.

**Definition 4.1.** *The logic* $\mathrm{AccLTL}^+$ *is the set of binding-positive formulas in* $\mathrm{AccLTL}(\mathrm{FO}^{\exists+}_{\mathrm{Acc}})$.

Note that in $\mathrm{AccLTL}^+$ we can describe the most basic dataflow constraint, the property of an access path being grounded: an access is grounded iff for every transition in a path, for every value that occurs in a binding, it occurs in some relation in the instance prior to the access:

$$\mathsf{G}\Big(\exists \vec{x}\ \mathrm{IsBind}_{\mathrm{AcM}}(x_1 \ldots x_m) \land$$
$$\bigwedge_{i \leq m} \bigvee_{R \in \mathrm{Sch}} \exists \vec{y}\, R(y_1 \ldots y_n) \land \bigvee_{j \leq n} y_j = x_i\Big)$$

Thus we can reduce satisfiability over grounded instances to satisfiability over all instances. Furthermore all the examples in the introduction are expressible in this fragment; we can express relevance of an access to a query as well as containment of queries under access patterns, restricting the paths to satisfy many data integrity, dataflow, and access ordering restrictions.

Our next main result is that this restriction suffices to give decidability:

**Theorem 4.2.** *Satisfiability of* $\mathrm{AccLTL}^+$ *is decidable in 3EXPTIME. The same is true for satisfiability over grounded instances and satisfiability over idempotent and exact accesses.*

We will show Theorem 4.2 by going through another specification formalism of interest in its own right, a natural automaton model for access paths. These are *Access-automata* (A-automata for short), which run over access paths, using a finite set of control states. At each transition $(I, (AcM, \bar{b}), I')$ of an access path the evolution function of the automaton tells what new states (if any) it can move to at the next position. The evolution function is a relational query that makes use of the binding, pre- and post- condition of the transition.

DEFINITION 4.3 (A-AUTOMATON). *Let* Sch *be a schema,* $Sch_{Acc}$ *the corresponding schema with accesses (as defined in Section 2), and $C$ a set of constants. An* Access-automaton *(A-automaton for short) over* $(Sch, C)$ *is a tuple* $(S, s_0, F, \delta)$ *where*
- $S$ *is a finite set of states,* $s_0 \in S$ *is an initial state,* $F \subseteq S$ *is a set of accepting states*
- $\delta$ *is a finite set of tuples of the form* $(s, \psi^- \wedge \psi^+, s')$ *where* $s, s'$ *are states,* $\psi^-$ *is a positive boolean combination of negated* $FO_{Acc}^{\exists+}$ *sentences that can not mention the predicate* IsBind*, while* $\psi^+$ *is a* $FO_{Acc}^{\exists+}$ *sentence; all these formulas can use constants in the given set* $C$.

**Semantics.** Let $A = (S, s_0, F, \delta)$ be an A-automaton and let $p$ be a path $t_1 \ldots t_n$ through the LTS associated with Sch, where $t_i = (I_i, (AcM_i, \bar{b}_i), I_{i+1})$. A *run* of $A$ on $p$ assigns to every $t_i$ a $\delta_i$ of the form $(s_i, \varphi_i, s_{i+1})$ in $\delta$ so that the relational structure $M(t_i)$ associated with $t_i$ satisfies $\varphi_i$. A run of $A$ is further said to be accepting iff its first state is initial and its last state is final. The language $L(A)$ accepted by an A-automaton $A$ is the set of access paths for which there is an accepting run. Note that an automaton only accepts access paths, which by definition must satisfy at least the property that for each $i$, $I_{i+1}$ extends $I_i$ solely by adding tuples to the relation of $AcM_i$, and all tuples added are consistent with the binding on the input positions of $AcM_i$. The definition of $L(A)$ can be further qualified to account for other sanity conditions (e.g. exactness).

A-automata are powerful enough to directly express relevance of an access in the presence of dataflow restrictions as well as disjointness constraints:

PROPOSITION 4.4. *Let $Q$ and $Q'$ be two positive queries, ACS a set of access methods, and $\Sigma$ a set of disjointness constraints. One can efficiently produce an A-automaton $A$ such that $Q$ is contained in $Q'$ under limited access patterns with disjointness constraints iff the language recognized by $A$ is empty. A similar statement holds for long-term relevance of an access to $Q$ under disjointness constraints.*

The proposition above can be extended to a general result stating that high-level logical specifications can be compiled into A-automata. We say that an A-automaton $A$ is equivalent to an AccLTL sentence $\varphi$ if the language of the $\varphi$ is the same as the language of $A$. The following result shows that each AccLTL$^+$ formula can be converted into an A-automaton.

LEMMA 4.5. *For each* AccLTL$^+$ *formula $\varphi$ there is an equivalent A-automaton of size exponential in the size of $\varphi$.*

We will show that emptiness of A-automata is decidable. Note that this decidability result together with Lemma 4.5 completes the proof of Theorem 4.2. Again, there are variants of the theorem for the various types of access, but we focus on the case of general accesses in the body of the paper.

THEOREM 4.6. *Emptiness of A-automata is decidable in 2EXPTIME. The same holds if accesses are restricted to be exact or idempotent.*

Notice that from Theorem 4.6 and Proposition 4.4 we get a 2EXPTIME upper bound for containment and long-term relevance. This improves on the prior known bounds [3, 5].

The proof uses a tight connection between A-automata and the containment problem for Datalog queries within positive first-order queries. This connection can also be exploited to give a corresponding lower bound:

THEOREM 4.7. *Emptiness of A-automata and satisfiability of* AccLTL$^+$ *are both 2EXPTIME-hard.*

### 4.1 Automata, Datalog, And Proof Sketch of Theorem 4.7

The proof of this result makes use of some new tools that we overview here. We reduce the emptiness problem for A-automata to the problem of whether a Datalog program is contained within a positive first-order query. Roughly speaking, we show that these automata can be captured by a conjunction of a Datalog query and the negation of a union of conjunctive queries. The reduction to this problem involves several stages, and the first step goes through a syntactic subclass of A-automata, called "progressive A-automata", defined below. We will show that the problem of testing emptiness of A-automata can be reduced to checking the emptiness of a bounded number of progressive A-automata.

**Progressive Automata.** In the following, given a boolean combination of $FO_{Acc}^{\exists+}$ formulas $\varphi$, we denote by $\tilde{\varphi}$ the formula $\exists \bar{x} \, \varphi'$ where $\varphi'$ is obtained from $\varphi$ by replacing each atom IsBind$_{AcM}(\bar{t})$ by $\bar{t} = \bar{x}$ and by replacing each predicate $R_{pre}$ by $R_{post}$. For a set $\Phi$ of sentences, we say that a formula is a *complete $\Phi$-type* if it is a conjunction that contains every formula of $\Phi$ either positively or negated. A formula is a "pure pre" (resp. "pure post") formula if it only mentions predicates of the form $R_{pre}$ (resp. $R_{post}$).

DEFINITION 4.8 (PROGRESSIVE A-AUTOMATON). *An A-automaton $A = (S, s_0, F, \delta)$ over $(Sch, C)$ is progressive if there is a pure pre formula $\Upsilon_{pre}(s_0)$ that does not use the predicate* IsBind$_{AcM}$*, a set of pure post $FO_{Acc}^{\exists+}$ sentences $\Phi$, and a function $\Upsilon_{post}$ mapping the states of $A$ to complete $\Phi$-types such that:*
1. *For any transition $(s, \varphi, s')$, if both* IsBind$_{AcM}(\bar{t})$ *and* IsBind$_{AcM'}(\bar{t}')$ *are atoms in $\varphi$, then* AcM = AcM'.
2. *For any transition $(s, \varphi, s')$, $\varphi$ implies $\Upsilon_{post}(s')$.*
3. *For any transition $(s_0, \varphi, s')$ that leaves the initial state, $\varphi$ implies $\Upsilon_{pre}(s_0)$.*
4. *For any transition $(s, \varphi, s')$ for which $s$ and $s'$ are in the same strongly connected component, $\Upsilon_{post}(s)$ is equivalent to $\Upsilon_{post}(s')$; also $\Upsilon_{post}(s')$ implies $\tilde{\varphi}$.*
5. *The maximal strongly connected components of $A$ form a sequence $C_1, \ldots, C_h$. That is, for each $i < h$, there is exactly one transition $(s, \varphi, s')$ such that $s \in C_i$ and $s' \in C_{i+1}$. For such a transition that connects two maximal strongly connected components, all atoms of the form* IsBind$_{AcM}(\bar{t})$ *must not contain variables; that is, $\bar{t}$ must be a sequence of constants.*
6. *The initial state is in $C_1$ and all accepting states are in $C_h$.*

We will call $h$ the *height* of $A$. A-automata correspond, up to emptiness, to unions of progressive automata.

LEMMA 4.9. *For every A-automaton $A$, there are progressive A-automata $A_1, \ldots, A_n$, such that, for each $i \leq n$, the size of $A_i$ is polynomial in the size of $A$, $n$ is exponential in the size of $A$, and $L(A)$ is empty iff $L(A_1) \cup \ldots \cup L(A_n)$ is empty.*

**From progressive A-automata to containment of Datalog in Positive Queries.** We now proceed to show that emptiness of progressive A-automata is decidable. Together with Lemma 4.9 this implies the decidability of (general) A-automata. This will involve reducing the emptiness of a progressive A-automaton to the problem of whether a Datalog program is contained in a positive first order logic sentence.

Recall that a Datalog program is defined with respect to two database schemas, called the *extensional schema* and the *intensional schema*. A *Datalog program* $\mathcal{P}$ is a finite set of rules of the form "head : – body" where head is an atomic formula $R(\bar{x})$ with a relation symbol $R$ in the intensional schema, and where body is a conjunctive query that can use relation symbols from the intensional and the extensional schema. Each Datalog program $\mathcal{P}$ contains a distinguished *goal predicate* $Q$. We use the standard notions of the least fixedpoint of a Datalog program $\mathcal{P}$ on a database $D$ (see [1]), and we denote this fixedpoint by $\mathcal{P}(D)$. We say that a Datalog program $\mathcal{P}$ *accepts* a database $D$ if the goal predicate of $\mathcal{P}$ is not empty in $\mathcal{P}(D)$.

LEMMA 4.10. *Let $A$ be a progressive A-automaton. Then there exists a Datalog program $\mathcal{P}_A$ and a positive first order logic sentence $\mathcal{P}'_A$ such that $L(A)$ is not empty iff $\mathcal{P}_A$ is not contained in $\mathcal{P}'_A$. One can construct these in polynomial time in the size of $A$.*

The proof of this lemma is itself quite involved. The basic idea of this proof is that $\mathcal{P}_A$ enforces the positive constraints of $A$ while $\mathcal{P}'_A$ enforces the negative constraints. Recall that in a progressive automaton, the evolution is in a fixed number of stages, based on the number of subqueries satisfied. A stage represents a strongly connected component of the automaton. The extensional database $D$ will have predicates BackgroundR$_i$ representing the part of relation $R$ that becomes visible to $A$ at the end of each stage $i$, along with predicates IntBackgroundR$_i$ representing the data that becomes visible when crossing from one stage to the next. The important intensional predicates ViewR$_i$ will represent intermediate stages of the predicates BackgroundR$_i$ within the evolution of each stage. The Datalog program $\mathcal{P}_A$ will have rules corresponding to the evolution of ViewR$_i$ by adding tuples from BackgroundR$_i$. To ensure that the tuples correspond to some valid binding, $\mathcal{P}_A$ will have rules guaranteeing that only tuples that satisfy the appropriate formulas can be added to ViewR$_i$. We can do this with a Datalog program by adding appropriate intermediate relations, exploiting the fact that the constraints on the guards are positive, and hence represented in non-recursive Datalog.

The role of the positive query $\mathcal{P}'_A$ is twofold: First, $\mathcal{P}'_A$ will enforce the negated conjunctive queries in the transitions – in particular, $\mathcal{P}'_A$ will contain constraints on the relations BackgroundR$_i$ and IntBackgroundR$_i$ that enforce that these only contain tuples that satisfy these negated constraints. In this way, whenever the Datalog program adds tuples to the intensional relations, these tuples are guaranteed to satisfy the corresponding negative constraints. The second purpose of $\mathcal{P}'_A$ is to enforce that for each $i$, only one relation among the IntBackgroundR$_i$ is non-empty. This is important, as these relations contain the tuples that the Datalog program might add when simulating the automaton transitioning from one strongly connected component to the next. On such a transition an A-automaton can only perform one access, and hence the Datalog program should only be able to add tuples from one relation IntBackgroundR$_i$ into ViewR$_i$.

In the proof that our construction is correct, we show that the Datalog program $\mathcal{P}_A$ can be decomposed into subprograms $\mathcal{P}_1, \ldots, \mathcal{P}_h$ that correspond to the decomposition of the A-automaton into strongly connected components $C_1, \ldots, C_h$ in the following sense: Whenever an A-automaton has a run that ends in its strongly connected component $C_i$, $i \leq h$ then the subprogram $\mathcal{P}_1 \cup \ldots \cup \mathcal{P}_i$ of $\mathcal{P}$ adds tuples to the intensional database that correspond in a certain way to the tuples that $A$ has obtained using accesses.

**Completion of the proof of Theorem 4.6.** Let us review what we have accomplished thus far: we have reduced questions about our logic to non-emptiness of the automata, and non-emptiness of an automaton we have reduced to determining whether a Datalog program is contained in a positive query. To complete the proof of Theorem 4.6 we need the following new result, that generalizes a theorem of Chaudhuri and Vardi [7]:

PROPOSITION 4.11. *The containment problem of a Datalog program $P$ in a positive first-order sentence $\varphi$, where both $P$ and $\varphi$ may make use of constants, is in $2EXPTIME$.*

The proof of this result is in the appendix. Theorem 4.6 follows from the proposition and the reduction given earlier.

## 4.2 Restricted Binding Predicates And Reduction To Propositional LTL

We now look for path query languages where the satisfiability problem has lower complexity. We will do this by giving up the ability to talk about the exact dataflow from data instances to bindings. This will allow us to get verification algorithms based on reduction to standard Propositional Linear Temporal Logic verification, a well-studied problem for which many tools are available [8].

For a relational schema Sch, we define the vocabulary Sch$_{0-\text{Acc}}$ as in Sch$_{\text{Acc}}$ but instead of the $n$-ary predicates IsBind$_{\text{AcM}}$, we have only a 0-ary predicate IsBind$_{\text{AcM}}$. A transition $t_i = (I_i, (\text{AcM}_i, \bar{b}), I_{i+1})$ is now associated with the relational structure $M'(t_i)$ in which $S_{pre}, S_{post}$ are interpreted as before, and IsBind$_{\text{AcM}}()$ holds exactly if AcM = AcM$_i$. We will now consider AccLTL(FO$_{0-\text{Acc}}^{\exists +}$), in which the first-order formulas use only Sch$_{0-\text{Acc}}$. That is, in the logic we can refer to which access was performed, but can not express anything about the bindings used.

Going back to Example 2.2 and 2.3 we say that the basic relevance properties are in this language, *provided* that we do not impose any dataflow restrictions – including any restrictions that access paths are grounded. On the other hand, we can still impose the access order restrictions of Example 2.3. We now see that by curtailing the expressiveness, the complexity goes down significantly.

THEOREM 4.12. *Satisfiability of an* $\text{AccLTL}(\text{FO}_{0-\text{Acc}}^{\exists+})$ *formula (over all access paths) is* PSPACE-*complete. The same holds if particular access methods must be exact or idempotent.*

PROOF. The PSPACE-hardness of our problem comes from the PSPACE-hardness of the satisfiability problem of a LTL formula over finite words [13]. The upper bound is proven by bounding the size of the underlying data, and then applying results about propositional LTL.

We now prove the upper bound, focusing on the case of general access paths. Let Sch be a schema, and $\varphi$ be a formula of $\text{AccLTL}(\text{FO}_{0-\text{Acc}}^{\exists+})$. First, we demonstrate that if there exists an access path that satisfies $\varphi$ then there exists one where the size of each instance is bounded by a polynomial function in the sizes of $\varphi$ and Sch.

The key is the following "Boundedness Lemma":

LEMMA 4.13. *An* $\text{AccLTL}(\text{FO}_{0-\text{Acc}}^{\exists+})$ *formula $\varphi$ is satisfiable iff there exists a path $\rho$ which satisfies the following properties: 1. The instances in $\rho$ have sizes bounded by a polynomial function in the sizes of $\varphi$, and* Sch. *2. The set of bindings used in $\rho$ has size bounded by a polynomial function in the sizes of $\varphi$*

PROOF. Let some $\varphi$ be given. Suppose that $\varphi$ is satisfiable. Then there exists a path $\rho$ that satisfies $\varphi$. We define the *positive sentences* of $\varphi$ to be the maximal subsentences of $\varphi$ that belong to $\text{FO}_{0-\text{Acc}}^{\exists+}$. Consider the following rewrite rules: for each AcM $\in$ Sch we replace the formula $\text{IsBind}_{\text{AcM}} \wedge \psi$, where $\text{IsBind}_{\text{AcM}}$ is a predicate, by the formula $\psi$. We also replace the formula $\text{IsBind}_{\text{AcM}} \vee \psi$ where $\text{IsBind}_{\text{AcM}}$ is a predicate by the formula $\psi$. We denote by $Q_f(\varphi)$ the set of $\text{FO}_{0-\text{Acc}}^{\exists+}$ sentences that have been obtained from a positive sentence of $\varphi$ by inductively applying the above rules until there are no more occurrences of predicates $\text{IsBind}_{\text{AcM}}$ in the result.

Let $\{q_1, \ldots, q_m\}$ be the set of sentences appearing in $Q_f(\varphi)$ that are satisfied by the last instance $I_n$. Let $\rho_{i_1}, \ldots, \rho_{i_m}$ be the set of transitions in the path $\rho$ such that $\rho_{i_j}$ is the minimal transition in $\rho$ that satisfies $q_j$. Let $h_j$ be a homomorphism from $q_j$ to $\rho_{i_j}$. We let $(I_{f-1}, AC_f, I_f)$ be the last transition in $\rho$. Let $I'_f$ be the minimal subinstance of $I_f$ such that for all $i$ $h_i(q_i) \subseteq (I'_f)^{pre} \cup (I'_f)^{post}$, where for any instance I of the original schema, $I^{pre}$ is obtained from I by interpreting relations $R_{\text{pre}}$ by the interpretation of $R$ in $I$, while $I^{post}$ is obtained from I by interpreting relations $R_{\text{post}}$ by the interpretation of $R$ in $I$.

Since we only need to consider witnesses to positive queries, it is easy to check that $I'_f$ can be constructed and has size polynomial in the sizes of $\varphi$ and Sch. We can thus construct a path $\rho'$ that contains the intersection of the instances of $\rho$ with the instance I'. $\rho'$ satisfies $\varphi$, and the size of the instances of $\rho'$ are bounded by a polynomial function in the size of $\varphi$ and Sch.

We now restrict the bindings used in $\rho'$. Let $p$ be a path. An access $(\text{AcM}_i, \vec{b}_i)$ is *necessary* for $p$ if new tuples are returned by it (i.e. tuples not in the previous instance within $p$), and *unnecessary* otherwise. Note that if we have a path and we change the binding on some unnecessary access to anything of the appropriate arity, while returning emptyset, then it is still a valid access path.

So without loss of generality, we can arrange that the set of bindings used in $\rho'$ consists of the necessary accesses in $\rho'$ plus a single binding for each access method, used in place of every unnecessary access on that method. Therefore the set of bindings is a set of tuples having size bounded by a polynomial function in the sizes of $\varphi$ and the schema. □

Given the lemma, we can now apply the following algorithm which is easily seen to be in NPSPACE:
1. First, we guess a finite sequence of instances $I_1 \ldots I_n$ and a sequence of accesses $A$, each of polynomial size (with the polynomial given by Lemma 4.13). In the remaining steps, we will check whether there is a witness path using the bindings of these accesses and only these instances.
2. We translate the $\text{AccLTL}(\text{FO}_{0-\text{Acc}}^{\exists+})$ formula $\varphi$ into an ordinary LTL formula $\overline{\varphi}$ in a propositional alphabet that encodes information about which of the instances and bindings are used. This formula will be constructed so that it is satisfiable over words iff $\varphi$ is satisfiable.
3. Then, we apply any PSPACE algorithm for LTL satisfiability of $\overline{\varphi}$ over finite words.

We now explain in more detail the translation to ordinary LTL that is the key step in the high-level algorithm above. Fix a sequence $s = I_1 \ldots I_n$ of distinct instances as well as a sequence of accesses $A$, both of polynomial size. We denote by $B$, the union of the set of bindings used in $A$ and the set $\cup_{\text{AcM}}\{b_{\text{AcM}}\}$ where $b_{\text{AcM}}$ is a binding of AcM using some values appearing in $B$.

We associate propositions with transitions of any of the following forms:

- Transitions of form $(I_i, (\text{AcM}, \vec{b}), I_i)$ where $\vec{b}$ is in $B$ and compatible with AcM.

- Transitions of form $(I_i, A_i, I_{i+1})$

The set of transitions of the above forms is denoted $T(I, B)$. For each $i$, we denote by $T(i)$ the set of transitions of the form $(I_i, (\text{AcM}, \vec{b}), I_i)$. For each $i$, we denote by $t_{i,\to}$ the transition $(I_i, A(i), I_{i+1})$. For each $i$, we denote by $P(i)$ the set of propositions associated with the transitions of the form $(I_i, (\text{AcM}, \vec{b}), I_i)$. For each $i$, we denote by $p_{i,\to}$ the proposition associated with the transition $(I_i, A_i, I_{i+1})$. The set of all such propositions is denoted $\Sigma$. The words described by $\overline{\varphi}$ are over alphabet $2^\Sigma$. Intuitively, each letter of a word would be used to describe a transition $(I, (\text{AcM}, \vec{b}), I')$.

We now describe the construction of $\overline{\varphi}$.

First, we describe some "sanity axioms" stating that a run associated with $\overline{\varphi}$ really corresponds to some access path. This requires:
- Every position has exactly one proposition of $\Sigma$.
- The order of the instances in $s$ is respected. This is expressed by the formula:

$$\bigwedge_{i, p \in P(i)} \mathsf{G}\Big(p \Rightarrow \Big(\bigvee_{p' \in P(i)} p' \mathsf{U} \, p_{i,\to}\Big)\Big) \wedge$$
$$\bigwedge_i \Big(p_{i,\to} \Rightarrow \mathsf{X}\Big(\bigvee_{p'' \in P(i+1)} p'' \vee p_{i+1,\to}\Big)\Big) \wedge$$
$$\Big(\bigvee_{p''' \in P(0)} p''' \vee p_{0,\to}\Big)$$

Next we rewrite $\varphi$ to $\overline{\varphi}$ by replacing each positive sentence $q$ of $\varphi$ by the union over of $p \in \Sigma$ over all the previous transitions that satisfy it.

We claim that the $\overline{\varphi}$ is satisfiable over ordinary words iff $\varphi$ is satisfiable over access paths that conform to the sequence $s$ and the bindings in $B$. The direction from right to left requires taking an access path and performing the obvious propositional abstraction. In the other direction, we take a propositional word $w_1 \ldots w_n$ satisfying $\overline{\varphi}$. The first sanity axiom implies that exactly one transition proposition $p$ is associated with $w_i$. The second sanity axiom implies that the instance reached in the transition associated with $w(i)$ is the same as the initial instance of the transition associated with $w(i+1)$. One can check that this gives the required access path for $\varphi$.

**Restricting LTL operators.** Let $\text{LTL}_X$ be the subset of LTL that only uses the temporal operator $\mathsf{X}$. We denote by $\text{AccLTL}(X)(\text{FO}_{0-\text{Acc}}^{\exists+})$ the corresponding sublanguage of $\text{AccLTL}(\text{FO}_{\text{Acc}}^{\exists+})$.

$\text{AccLTL}(X)(\text{FO}_{0-\text{Acc}}^{\exists+})$ is extremely limited in expressiveness, since it can only talk about paths of some fixed length. However, there are properties for which such small paths are sufficient. Consider Example 2.3. It is easy to see that $Q$ is LTR over all accesses iff it is LTR over access paths of size $|Q|$ – a counter example to long-term relevance has only polynomially length. But LTR over small paths can be expressed in $\text{AccLTL}(X)(\text{FO}_{0-\text{Acc}}^{\exists+})$. Thus $\text{AccLTL}(X)(\text{FO}_{0-\text{Acc}}^{\exists+})$ is sufficient to tell whether an access might have an impact on answering a query, but without taking into account of even the most basic dataflow restriction on paths.

THEOREM 4.14. *Satisfiability of* $\text{AccLTL}(X)(\text{FO}_{0-\text{Acc}}^{\exists+})$ *is $\Sigma_2^P$-complete, even when certain accesses are restricted to be exact or idempotent.*

*Hardness.* Non-containment of positive relational queries, where positions can be restricted to have finite (i.e. enum) datatypes can be reduced to the unsatisfiability problem of either language – this problem is known to be $\Pi_2^P$-hard.

*Upper-Bound.* Let an $\text{AccLTL}(X)(\text{FO}_{0-\text{Acc}}^{\exists+})$ formula $\varphi$ be given. We first note that Lemma 4.13 also holds for the logic $\text{AccLTL}(X)(\text{FO}_{0-\text{Acc}}^{\exists+})$. Using this we can reduce to the language propositional $\text{LTL}_X$, which has a satisfiability problem in NP. In the reduction we will again guess a small number of small instances and bindings, and we will also guess which positive queries of $\varphi$ will be true – this guess will then be verified via a sequence of NP (for queries guessed to be true) and co-NP (for queries guessed to be false) subroutines. We can then rewrite the original formula $\varphi$ to an $\text{LTL}_X$ formula that is satisfiable iff $\varphi$ is satisfiable on a sequence based on the guessed instances and bindings.

## 5. EXTENSIONS AND LIMITS

We look at the impact of two natural extensions on our decidability results: allowing inequalities and branching formulas.

### 5.1 Extension To Inequalities

Our results on decidable fragments did not use inequalities, and inequalities are useful for expressing data integrity constraints. The most obvious example involves keys and functional dependencies, as discussed in Example 2.4.

By making a straightforward modification of the proofs without inequalities, we can see that inequalities add nothing to the complexity of $\text{AccLTL}(\text{FO}_{0-\text{Acc}}^{\exists+})$ and its sublanguages.

THEOREM 5.1. *Letting* $\text{FO}_{0-\text{Acc}}^{\exists+,\neq}$ *be the language of positive queries with inequalities over the restricted vocabulary with only the 0-ary predicates* $\text{IsBind}_{\text{AcM}}$, *we have that*
- *satisfiability of* $\text{AccLTL}(\text{FO}_{0-\text{Acc}}^{\exists+,\neq})$ *is in* PSPACE *(and hence* PSPACE*-complete by Theorem 4.12)*
- *satisfiability of* $\text{AccLTL}(X)(\text{FO}_{0-\text{Acc}}^{\exists+,\neq})$ *is in* $\Sigma_2^P$ *(hence $\Sigma_2^P$-complete by 4.14)*

Using the language above, one can express relevance or containment in the presence of functional dependencies, access order constraints, and disjointness constraints, but not dataflow constraints.

For the language $\text{AccLTL}^+$, shown decidable in Theorem 4.2, inequalities make a dramatic difference. The proof of the theorem below shows that we cannot capture both dataflow restrictions like groundedness along with rich integrity constraints such as functional dependencies, while retaining decidability. The proof also shows that many extensions of $\text{AccLTL}^+$ with aggregation – basically, any that are expressive enough to capture FDs – will be undecidable.

THEOREM 5.2. *For binding-positive* $\text{AccLTL}(\text{FO}_{\text{Acc}}^{\exists+,\neq})$, *satisfiability is undecidable.*

PROOF. Again we reduce the problem of implication of functional dependencies (fds) and inclusion dependencies (ids) for relational databases to the problem of the unsatisfiability of a $\text{AccLTL}^+$ extended with inequalities.

Let $\Gamma$ be a set of inclusion and functional dependencies, and $\sigma$ be a functional dependency over Sch.

The approach to the reduction is similar to that in Theorem 3.1. We will make iterative accesses to a successor relation of a total order over the tuples. We will also access relations $\text{Beg}(R)$ and $\text{End}(R)$, and verify that they contain the first and the last tuples of relation $R$ according to the order. While iterating through the relations according to the successor relation, the satisfaction of the different fd's and id's and the failure of $\sigma$ are verified. The satisfaction and failure of fd's can be reduced to the satisfaction of a boolean combination of conjunctive queries with inequalities – the successor relation is not needed. The satisfaction of an inclusion dependency $id$ whose source is a relation $R$ is where we use the successor relation, and the iteration technique of Theorem 3.1. Again, it is easy to check an inclusion dependency for a source relation consisting of only a single tuple, since this requires only existential quantification. We verify an id on source relation $R$ by checking for witnesses for one tuple in the source of the dependency at a time, iterating on the tuples according to the successor relation. We will use a new predicate $\text{CheckIncDep}(id)$ whose arity is the arity of $R$. $\text{CheckIncDep}(id)(\vec{t})$ holding at some instance indicates that $\vec{t}$ has been verified to satisfy the inclusion dependency $id$. This will be done in a "loop" (an "until" in the logic) in which we look for a tuple $\vec{t}$ whose predecessor in the order satisfies $\text{CheckIncDep}(id)$, and which satisfies the inclusion dependency; when we find such a tuple, we perform an access to $\text{CheckIncDep}(id)$ on it. At the end of this "loop", we check that the final tuple in the ordering satisfies $\text{CheckIncDep}(id)$. □

The reader may want to look at Figure 2 for a view to how the languages with inequalities relate to the languages defined previously.

## 5.2 Branching Time Formulas

Thus far we have discussed only linear time properties of the LTS of a schema with access relations. What about branching time logics, which can consider the relationship of multiple paths? For example, a branching time logic could express that we have reached a point where no further information about boolean query $Q$ can be obtained without guessing values to enter into forms – e.g. there are possible worlds consistent with the known facts where $Q$ is true and also consistent worlds where $Q$ is false, but the truth of $Q$ can not be revealed by any further sequence of grounded accesses. Unfortunately, we will show that even very limited branching time expressiveness leads to undecidability.

Let $L$ be a fragment of first-order logic over the smallest vocabulary we have considered thus far: two copies $S_{\text{pre}}, S_{\text{post}}$ of each relation symbol $S$ and the proposition $\text{IsBind}_{\text{AcM}}$.

We will consider a small fragment of branching time logic built up from $L$-formulas, analogously to the way we built up AccLTL formulas over sentences of $L$ in the linear time logic. Traditional branching time logic allows the combination of path quantification with modal operators. In our setting we will consider a very simple kind of branching, which looks ahead only one step – we will refer to it as $\text{CTL}_{EX}(L)$, but instead of CTL we might as easily have said "basic modal logic" or Hennessy-Milner Logic [13], since we only need the power of the most basic existential modality to get undecidability. $\text{CTL}_{EX}(L)$ has the rules: every $L$ sentence is a formula, boolean combinations of formulas are formulas, and if $\varphi$ is a formula then $\text{EX}\varphi$ (in modal logic notation, $\diamond\varphi$) is a formula.

The semantics is defined as a relation $(S, t) \vDash \varphi$, where $t$ is a transition $(I, AC, I')$ in the labelled transition system $S$ associated with a schema Sch. When $\varphi$ is an $L$ formula, this holds iff the relational structure associated to $t$, $M'(t)$, satisfies $\varphi$ in the usual sense of first-order logic. The semantics of boolean operators is the usual one. Finally, $(S, t) \vDash \text{EX}\varphi$ iff there is a successor $t'$ of $t$ such that $(S, t') \vDash \varphi$. Note that instead of referring to CTL here, we could have used basic modal logic or Hennessy-Milner Logic. Note that Deutsch et. al. [12] have shown undecidability for some branching time logics over LTS's associated with a similar model of relational transducers – but in their case the logics (e.g. Theorem 4.14 of [12]) allow one to describe properties of the input (analogous to our larger signature $\text{Sch}_{\text{Acc}}$), while here we can only describe the access propositionally.

We show that even this restricted logic is undecidable, even when the base formulas are existential.

THEOREM 5.3. *Satisfiability of $\text{CTL}_{EX}(\text{FO}_{0-\text{Acc}}^{\exists+})$ is undecidable*

PROOF. We reduce from the problem of implication of a functional dependency (FD) from a set of functional dependencies and inclusion dependencies (IDs) for relational databases. This is known to be undecidable [6].

Let $\Gamma$ be a set of inclusion and functional dependencies over a relational schema Sch and $\sigma$ an FD. For simplicity, we will assume all positions in the schema have the same type (say, integer type). We will first extend Sch with additional relations, along with access patterns.

For each relation $R$ of Sch, we have an access method $\text{Fill}_R$ on $R$ with no inputs. Thus each access $(\text{Fill}_R, \varnothing)$ returns an essentially random configuration of $R$. We also have additional relations $\text{Chk}^{\text{FD}}(R)$, having twice the arity of $R$ and $\text{CheckIncDep}(R)$ having the same arity as $R$. We have boolean access methods on all of these additional relations – that is, methods where all positions are in the input.

Our reduction will create a formula $\psi(\Gamma, \sigma)$ of the form:

$$\text{EX}\Big(\text{Fill}_{R_1} \wedge \text{EX}(\cdots \wedge \text{EX}(\text{Fill}_{R_n} \bigwedge_{\text{fd}\in\Gamma} \varphi_{\text{fd}} \wedge \bigwedge_{\text{id}\in\Gamma} \varphi_{\text{id}} \wedge \varphi_{\neg\sigma}))\Big)$$

where $\varphi_{\text{fd}}, \varphi_{\text{id}}$, and $\varphi_{\neg\sigma}$ will be defined below, but we explain their mission now. For each functional dependency $\text{fd} \in \Gamma$, the formula $\varphi_{\text{fd}}$ will hold on a transition $t = (I, AC, I')$ exactly when fd holds on the restriction of $I'$ to the schema predicates from Sch, and similarly for $\varphi_{\text{id}}$. The formula $\varphi_{\neg\sigma}$ checks that $I'$ does not satisfy the functional dependency $\sigma$. Thus this formula will imply that the configuration is a witness showing that $\Gamma$ does not imply $\sigma$.

We now explain how the different formulas are built. Let $\text{fd} = R : P \to p$ where $P$ are positions of relation $R$ and $p$ is a position of $R$. The formula $\varphi_{\text{fd}}$ will be:

$$\text{AX}\Big(\exists\vec{x}\vec{y} \ \text{Chk}^{\text{FD}}(R)_{\text{post}}(\vec{x}, \vec{y}) \wedge$$
$$\bigwedge_{i\in P} x_i = y_i \wedge R_{\text{post}}(\vec{x}) \wedge R_{\text{post}}(\vec{y})$$
$$\Rightarrow \exists\vec{x}'\vec{y} \ \text{Chk}^{\text{FD}}(R)_{\text{post}}(\vec{x}', \vec{y}') \wedge x'_p = y'_p\Big)$$

Here we use the derived "box" modality $\text{AX}\varphi = \neg\text{EX}\neg\varphi$. Note that $\varphi_{\text{fd}}$ occurs in formula $\psi(\Gamma, \sigma)$ in a context where we know that only accesses to $R_i$ have been done – hence only in contexts where $\text{Chk}^{\text{FD}}(R)$ must be empty. Since the only access methods for the relations $\text{Chk}^{\text{FD}}(R)$ are boolean, this means that after one transition we can have at most one tuple in $\text{Chk}^{\text{FD}}(R)_{\text{post}}(\vec{x}, \vec{y})$. Thus doing a modality $\text{AX}$ followed by a test that $\text{Chk}^{\text{FD}}(R)(\vec{x}, \vec{y}) \wedge R_{\text{post}}(\vec{x}) \wedge R_{\text{post}}(\vec{y})$ holds amounts to testing an arbitrary pair $\vec{x}, \vec{y}$ satisfying $R$ prior to the access. The formula thus asserts that for any such pair of tuples in $R$, if they agree on all positions in the source of the FD, they agree on the target of the FD.

We can use a similar trick with the formula $\varphi_{\neg\sigma}$:

$$\text{EX}\Big(\exists\vec{x}\vec{y} \ \text{Chk}^{\text{FD}}(R)_{\text{post}}(\vec{x}, \vec{y}) \wedge \bigwedge_{i\in P} x_i = y_i \wedge$$
$$R_{\text{post}}(\vec{x}) \wedge R_{\text{post}}(\vec{y}) \wedge$$
$$\neg\exists\vec{x}'\vec{y} \ \text{Chk}^{\text{FD}}(R)_{\text{post}}(\vec{x}', \vec{y}') \wedge x'_p = y'_p\Big)$$

Now fix an id $R[A_1, \cdots, A_n] \subseteq S[B_1, \cdots, B_n]$, and we define $\varphi_{\text{id}}$ to be

$$\text{AX}\Big(\text{IsBind}_{\text{CheckIncDep}(R)} \wedge R_{\text{post}}(\vec{x}) \wedge$$
$$\exists\vec{x} \ \text{CheckIncDep}(R)_{\text{post}}(\vec{x}) \Rightarrow$$
$$\text{EX}\big(\text{IsBind}_{\text{CheckIncDep}(S)} \wedge \exists\vec{x} \ \text{CheckIncDep}(R)_{\text{post}}(\vec{x}) \wedge$$
$$\exists\vec{y}\text{CheckIncDep}(S)_{\text{post}}(\vec{y}) \wedge \bigwedge_{i\le n} x_{A_i} = y_{B_i}\big)\Big)$$

This states that whenever we do a "test access" that returns an element of $R$, there is some access we can do immediately afterwards in the LTS that reveals a matching tuple in $S$. As in the case of $\varphi_{\text{fd}}$ above, the accesses we perform are boolean, and hence cannot be creating any new elements of $S$ – thus the revealed match must have been in the configuration prior to the access. □

Table 1: Complexity and application examples for path specifications.

| Language | Complexity | DjC | FD | DF | AccOr |
|---|---|---|---|---|---|
| AccLTL($FO_{Acc}^{\exists+,\neq}$) | undecidable | Yes | Yes | Yes | Yes. |
| AccLTL($FO_{Acc}^{\exists+}$) | undecidable | Yes | No | Yes | Yes |
| AccLTL$^+$ | in 3EXPTIME | Yes | No | Yes | Yes |
| A-automata | 2EXPTIME-compl. | Yes | No | Yes | Yes |
| AccLTL($FO_{0-Acc}^{\exists+}$) | PSPACE-compl. | Yes | No | No | Yes |
| AccLTL($FO_{0-Acc}^{\exists+,\neq}$) | PSPACE-compl. | Yes | Yes | No | Yes |
| AccLTL(X)($FO_{0-Acc}^{\exists+,\neq}$) | $\Sigma_2^P$-compl. | Yes | Yes | No | No |

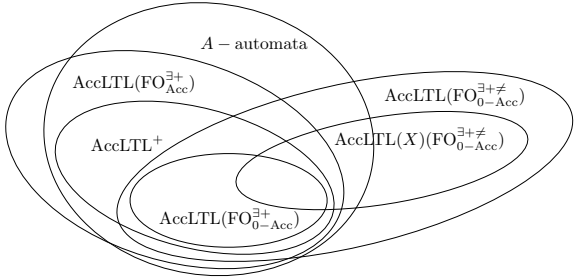

Figure 2: Inclusions between language classes.

## 6. CONCLUSIONS AND RELATED WORK

In this work we introduced the notion of querying the access paths that are allowed by a schema. We presented decidable specification languages for doing this, and gave undecidability results showing several limits of such languages. Figure 2 shows the inclusions of the languages considered in the paper, excluding those for branching time. All of the containments shown in the diagram are straightforward. The containment of $FO_{0-Acc}^{\exists+}$ in AccLTL$^+$ does require one to deal with the fact that $FO_{0-Acc}^{\exists+}$ sentences are not required to be binding-positive. The inclusion follows by first rewriting negated 0-ary IsBind$_{AcM}$ predicates using the rule IsBind$_{AcM}$ = $\bigvee_{AcM' \neq AcM}$ IsBind$_{AcM'}$, then replacing the 0-ary predicate by existentially-quantified $n$-ary predicates.

All the inclusions in the diagram also turn out to be strict. We omit the proofs for this, which use standard techniques: e.g. A-automata can express parity conditions on the length of paths, which first-order languages like AccLTL$^+$, or even AccLTL($FO_{Acc}^{\exists+}$), can not do.

Table 1 shows the complexity of satisfiability for each specification formalism, along with application examples. DjC indicates that the language can express relevance of an access in the presence of disjointness constraints, while FD, DF, AccOr refer to functional dependencies, dataflow restrictions, and access order restrictions, respectively.

Our work leaves open a number of questions concerning the logics we study – for example, we leave open the exact complexity of AccLTL$^+$, which lies between double- and triple- exponential time. We also do not have tight bounds for our more restricted fragments (e.g. with only the 0-ary version of IsBind$_{AcM}$) in the important case of grounded access paths.

Although this is, to the best of our knowledge, the first work on languages for describing access paths through a schema with binding patterns, there is a strong formal connection to work on verifying data-driven services, as well as other work in the area of hidden Web querying. We review the closest connections below.

**Data-driven services.** Our work is closely related to a line of research on relational transducers and models for data-driven services, beginning with Abiteboul et. al.'s [2], and continuing through work of Spielmann [19], Deutsch, Su and Vianu (e.g. [12]), Fritz et. al. [14], and Deutsch et. al. [10]. All of these works deal with specification languages for transition systems in which transitions may involve the consuming of relational inputs from an external environment, the production of output tuples, and the modification of internal state (perhaps in the form of an additional relational store). In our application, we talk of accesses rather than inputs from an environment, with a response consisting of revealing a hidden database instance, rather than updating an internal store. But in the results of this paper, one can just as easily think of identifying the hidden Web database with an internal store, with the accesses being non-deterministic inserts into the store.

Nevertheless, the logics that arise naturally in our setting appear orthogonal to those studied in prior work. The initial Abiteboul et. al. paper [2] focused on "Spocus transducers" (semi-positive output and cumulative state) which take full relational inputs, with their internal relations only accumulating them. A direct comparison with our model is difficult, since we do not have a notion of "output" – but if we restrict Spocus transducers to boolean output and singleton inputs, they are not as powerful as our model, since in our case the internal state can be modified in non-trivial ways. [2] proves an undecidability result for an extension of Spocus transducers in which the inserted data is allowed to be a projection of the "input relations" (Prop. 3.1 of [2]). The technique applied is similar to that in Theorem 5.2, but projection is orthogonal to the update given by access methods. In our terms, this extension would amount to having the information added to the hidden database be a projection of the accessed relations. On the other hand, the addition of projection does not give the ability to model access methods, which restrict the input relations by requiring them to satisfy a selection criterion.

Later works [19, 12, 10, 14] deal with transducers that can delete as well as insert into their internal state. A key restriction is *input-guardedness*, which insures decidability [12] – input guardedness requires quantifications to be restricted to tuples generated from the environmental inputs. The analogous restriction in our setting would be to restrict

644

quantification to the bindings, which would be much weaker than the logics we consider. Thus our decidability and complexity results are not subsumed by these works. On the other hand, guarded quantification over relational inputs is not supported by our logics, and hence we do not claim to subsume results in these works. In addition, [10] allows a built-in linear order on the domain, which we do not consider for our largest logics. Later work by Damaggio et. al. considers even richer signatures, including arithmetic [9].

**Hidden Web querying.** Our work is directly inspired by previous results on static analysis of schemas with limited access patterns, a line of work tracing back (at least) to Ullman's work [20] and Rajaraman et. al. [18], continuing with Chang/Li's work in the early 2000s [16, 15] Ludäscher/Nash's and Deutsch et. al.'s work in the mid-2000's [17, 11] and Cali et. al. [5]. All of them deal in one way or another with what sequences can occur within a sequence with limited access patterns. For example, the question of whether a query can always be answered using exact grounded access paths – the focus of most of these works above – can be expressed as a property of the LTS. Exact complexity bounds for query answering derived from the works above. Containment under access patterns has also been studied, particularly in [5], which establishes a coN-EXPTIME upper bound for conjunctive queries. [3] proves a matching coNEXPTIME lower bound for containment for conjunctive queries, and a co-2NEXPTIME upper bound for positive queries. [3] also defines the notion of long-term relevance (LTR). They prove a $\Sigma_2^p$-completeness result for LTR over general access paths ("independent accesses", in their terminology) while providing a NEXPTIME-completeness result for conjunctive queries and a 2NEXPTIME bound for LTR of positive queries over grounded accesses paths ("dependent accesses").

Our work provides a general framework where we can express properties of access paths, including containment, LTR, their combinations, and their restrictions to constraints. By providing these within a boolean closed logic, we give a flexible means of combining properties that one wishes to verify. Our 2EXPTIME result for non-emptiness of $A$-automata gives a bound on containment under access patterns and long-term relevance, as mentioned in the discussion after Theorem 4.6. This is better than the prior bounds from [5, 3].

Note that [3] also makes some erroneous claims: 1. A co2NEXPTIME lower bound for containment of positive queries under access patterns, which is at odds (relative to complexity-theoretic hypothesis) with our 2EXPTIME upper bound 2. A coNEXPTIME upper bound for containment of UCQs under general access patterns. The proof given there only works for schemas with a single-access per relation, while in subsequent work, we have shown that the problem is 2EXPTIME hard if the single-access restriction is dropped.

## 7. ACKNOWLEDGMENTS

Benedikt and Bourhis are supported in part by EC FP7-ICT-233599 (*FoX*) and in part by EP/G004021/1 and by EP/H017690/1 of the Engineering and Physical Sciences Research Council UK.